\documentclass[final,3p,sort&compress,times,twocolumn]{elsarticle}

\usepackage{amssymb,amsmath,amsthm}
\journal{Physics Letters A}

\usepackage{hyperref}  
\usepackage{tabto}  
\usepackage[euler]{textgreek}
 \usepackage{ragged2e}
 \usepackage{graphicx}
\usepackage[table,xcdraw]{xcolor}
\begin{document}

\begin{frontmatter}

\title{Significance of Composition-Dependent Effects in Fifth-Force Searches}

\author[1,2]{Ephraim Fischbach\corref{cor1}}%
\ead{ephraim@purdue.edu}

\cortext[cor1]{Corresponding author}

\author[2]{John T. Gruenwald}
\author[3,1]{Dennis E. Krause}
\author[1]{Megan~H.~McDuffie}
\author[1]{Michael~J.~Mueterthies}
\author[4,5]{Carol~Y.~Scarlett}

\address[1]{Department of Physics and Astronomy, Purdue University, West Lafayette, IN 47907 USA}
\address[2]{Snare, Inc., West Lafayette, IN 47906, USA}
\address[3]{Department of Physics, Wabash College, Crawfordsville, IN, 47933 USA}
\address[4]{Department of Physics, Florida A\&M University, Tallahassee, FL 32307 USA}
\address[5]{Orise Fellow, Argonne National Laboratory, Lemont, IL 60439 USA}

\begin{abstract}
Indications of a possible composition-dependent fifth force, based on a reanalysis of the E\"{o}tv\"{o}s experiment, have not been supported by a number of modern experiments. Here, we argue  that searching for a composition-dependent fifth force necessarily requires data from experiments in which the acceleration differences of three or more independent pairs of test samples of varying composition are determined. We suggest that a new round of fifth-force experiments is called for, in each of which three or more different pairs of samples are compared. 
\end{abstract}

\end{frontmatter}

\section{Introduction}

Today it is often taken for granted that the effect of gravity on a body is independent of its composition, but this was not always  the case.  While Galileo recognized that the notion that heavier objects fell faster than lighter objects in a gravitational field was incorrect, his experiments involving rolling balls of different composition down inclined planes suffered from significant systematic effects. (See Ref.~\cite{Franklin Laymon} for a recent discussion of Galileo's gravity experiments.)  It is almost certain that any experiments attributed to him dropping different objects from the Tower of Pisa are either apocryphal, or could not have led to convincing conclusions due to air resistance.  However, he was able to show that two pendulums using test bodies of lead and cork maintained the same period despite the significant difference in masses.

A more extensive test of what is now called the  weak equivalence  principle (WEP) was conducted by Newton who used  pendulums with boxes filled with gold, silver, lead, glass, sand, common salt, wood, water and wheat.  He showed that the differences in their gravitational and inertial masses or equivalently,  the differences $\Delta a$ in their  accelerations were $\Delta a/g \lesssim 10^{-3}$, where $g$ is the local acceleration of gravity \cite{Franklin Laymon}. The pendulum technique was improved upon by Bessel \cite{Bessel} in 1832 using Fe, Zn, Pb, Ag, Au, Fe$_{2}$O$_{2}$, marble, clay, quartz,  and H$_{2}$O,  who found that $\Delta a/g \lesssim 2 \times 10^{-5}$ \cite{Potter}.  Later, in 1923,  Potter obtained $\Delta a/g \lesssim 3 \times 10^{-6}$  for  materials including lead, steel, ammonium fluoride, bismuth, paraffin wax, mahogany, and duralumin \cite{Franklin Laymon,Potter}.  

A significant leap for the test of the composition independence of gravity was achieved by a series of experiments between 1904 and 1908 by  E\"{o}tv\"{o}s, Pek\'{a}r, and Fekete (EPF)  using a torsion balance of their own design \cite{EPF,EPF Book 1,EPF Book 2}.  By comparing 11 different combinations of materials, they found $\Delta a/g \lesssim 10^{-9}$.  Since then, WEP tests have  become significantly more stringent using improved torsion balances \cite{Wagner}, free-fall experiments \cite{Niebauer,Cavasinni}, and, more recently, atom interferometers \cite{Asenbaum}, and an orbiting satellite (MICROSCOPE) \cite{Touboul PRL,Touboul CQG}.  A recent comprehensive review of tests of gravity and the WEP can be found in Ref.~\cite{Tino}.  

While the early experiments probing the WEP were focused on understanding gravity or testing general relativity, for which the WEP is a cornerstone,  this is not the case for more recent experiments.  There is now overwhelming evidence for the correctness of general relativity as {\em the} classical theory of gravity.   Instead, a major factor driving modern  tests of the WEP is the search for new forces coexisting with gravity motivated, in part,  by a reanalysis of the EPF experiment by Fischbach and colleagues  \cite{Fischbach PRL,AoP} which will be discussed in more detail in the next section.   It is now recognized that most extensions of the Standard Model of particle physics, including string theory, will lead to new composition-dependent forces.  Furthermore, most laboratory experiments searching for dark matter are hoping that dark matter will interact with ordinary matter via non-gravitational interactions, which likely require new physics.  Therefore, tests of the WEP remain at the forefront of fundamental physics centuries after Galileo and Newton.

The purpose of this paper is to re-examine the results of the EPF experiment, the most precise test of the WEP which used a wide variety of materials, in contrast to more recent experiments which typically use just a single pair of different materials.  Since a hallmark of a new force is its composition-dependence, we argue that new experiments searching for new forces should recall  earlier experiments testing gravity, and use a wider variety of materials.  This will allow one to  probe more general force couplings, and to reduce the possibility that the design of an experiment might inadvertently  diminish the sought-after signal.

\section{Reanalysis of the EPF Experiment}

Interest in the EPF experiment was revived in the  1980s following hints by experiments involving geophysical measurements and the $K^{0}$-$\bar{K}^{0}$ system of  the existence of a  new intermediate-range force coupling to baryon number or hypercharge \cite{Fischbach memoir,Franklin}.  The possibility of a new force coupling to baryon number had been  considered by Lee and Yang in 1955 \cite{Lee}.  They briefly suggested the EPF experiment as a test, since such a new force would cause an {\em apparent} violation of the WEP, given that this new composition-dependent force would coexist with gravity.  Fischbach and colleagues realized that evidence for a new  ``5th force'' coupling to baryon number with a range of order $ 10^{2}$~m might be found in  data from the  EPF experiment \cite{Fischbach memoir,Franklin}.  While more sensitive tests of the WEP since EPF had been conducted by a group led by Dicke \cite{Dicke}, and by Braginskii and Panov \cite{Braginskii}, these experiments compared accelerations toward the Sun, and so would not have been sensitive to a force with a range much shorter than the Earth-Sun separation.  The reanalysis of EPF data  by Fischbach et al.\  showed that the data were, in fact,  consistent with the presence of a new composition-dependent interaction proportional to the baryon numbers $B$ of the samples \cite{Fischbach PRL,AoP}.   

To explain the geophysical observations and the EPF data, Fischbach et al. proposed the existence  of a new force coexisting with gravity, such that the total potential energy of two point masses $m_{1}$ and $m_{2}$ separated by a distance $r$ was given by
  \begin{equation}
  V(r) = V_{N}(r) + \Delta V(r) = -\frac{Gm_{1}m_{2}}{r}\left(1 + \alpha e^{-r/\lambda}\right),
  \label{V}
  \end{equation}
  where $V_{N}(r)$ is the Newtonian gravitational potential, and $\Delta V(r)$ is a new Yukawa potential with range $\lambda$.   In Eq.~(\ref{V})    
  \begin{equation}
  \alpha = f^{2}\left(\frac{B_{1}}{m_{1}}\right)\left(\frac{B_{2}}{m_{2}}\right) = \frac{f^{2}}{m_{H}^{2}}\left(\frac{B_{1}}{\mu_{1}}\right)\left(\frac{B_{2}}{\mu_{2}}\right)
  \label{alpha}
  \end{equation}
is the strength of the new interaction relative to gravity.
Here $f$ is the unit of charge for the baryonic fifth force, and $\mu_{i}$ is the mass of the $i$th sample in units of atomic hydrogen, with $m_H = m(_1H^1) = 1.00782519(8)$u.  For two samples $i$ and $j$ compared in the EPF experiment, this model predicts that the acceleration difference between the samples and the Earth is
\begin{equation}
\Delta \kappa \equiv \frac{\Delta a}{g} = \frac{f^2\epsilon(R/\lambda)}{G m^2_H} \left(\frac{B_\oplus}{\mu_\oplus}\right)\left(\frac{B_i}{\mu_i} - \frac{B_j}{\mu_j}\right),
\label{Delta kappa 5}
\end{equation}
where we can approximate the baryon-to-mass ratio of the Earth, $B_\oplus/\mu_\oplus \simeq 1$, for present purposes. The function $\epsilon(R/\lambda)$ arises from the integration of the baryon number distribution of the Earth, the presumed source, here modeled as a sphere of radius $R$, under the assumption that the new baryonic force has a finite range $\lambda$. With $x = R/\lambda$, $\epsilon(x)$ is given by 
\begin{equation}
\epsilon(x) = \frac{3(1+x)}{x^3}e^{-x}\left[x\cosh(x) - \sinh(x)\right].
\end{equation} 

\begin{figure} [t]
\includegraphics[width=\columnwidth]{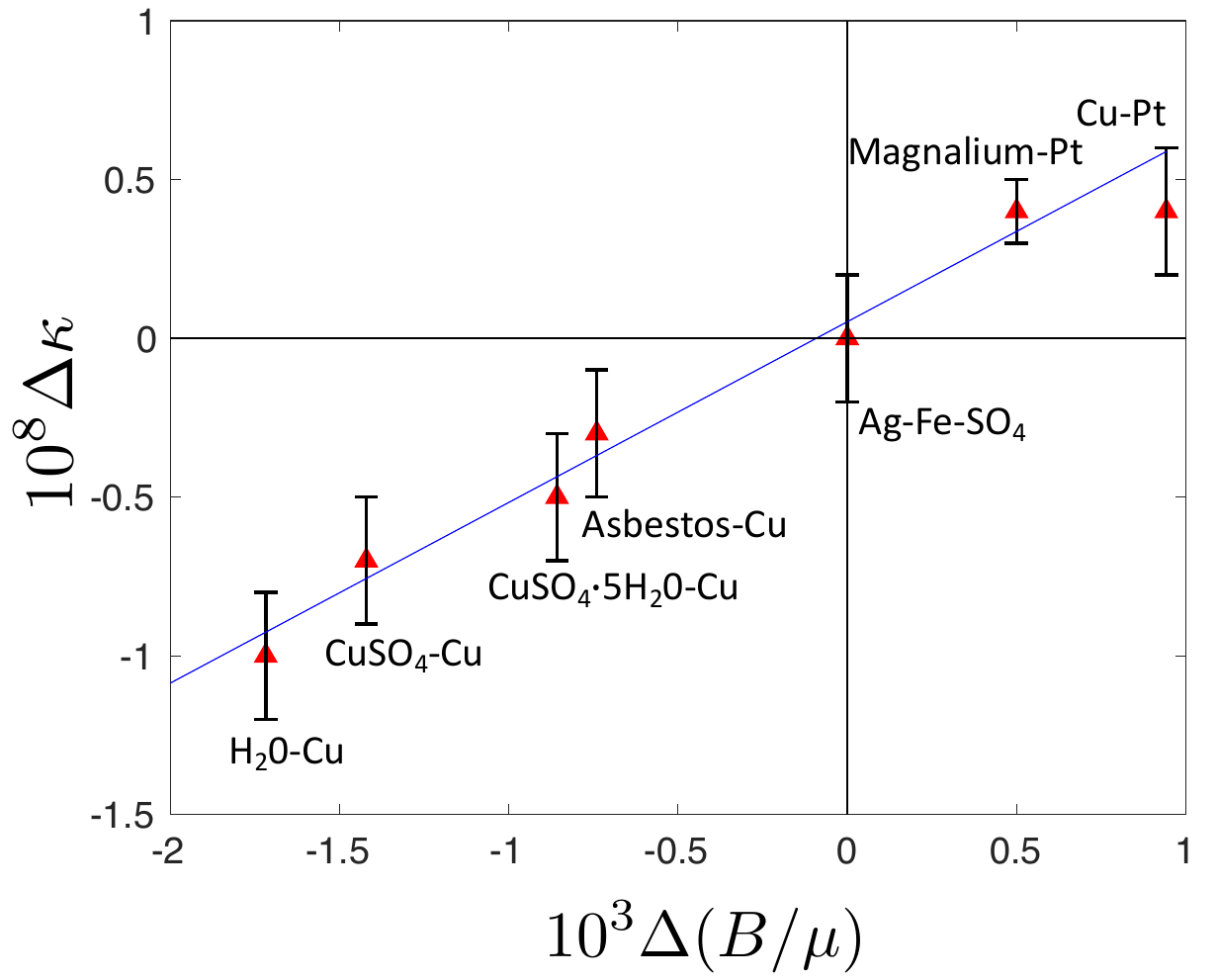}
\caption{The original EPF data with uncertainties for the seven materials in Table 1 of Ref.~\cite{Fischbach PRL}.  Note that only one data point, Ag-Fe-SO$_4$, is consistent with the null hypothesis at the $1\sigma$ level.}
\label{Original EPF graph}
\end{figure}

The results of the reanalysis of the EPF data with the proposed fifth force model are presented in Table~\ref{EPF data table} and Fig.~\ref{Original EPF graph}.  The data clearly show  that the data were consistent with the presence of a new composition-dependent interaction proportional to the baryon numbers $B$ of the samples \cite{Fischbach PRL,AoP}.   Although the data from the EPF experiment were interpreted by the authors themselves, as well as by subsequent researchers, as having given a null result consistent with conventional Newtonian gravity, it is interesting to note  that most of their data points are {\em not} consistent with the WEP at the $1\sigma$ level, which may have been responsible for the delay in the publication of the results \cite{Fischbach PoS}.  

Since 1986, numerous  searches for  a fifth force of the form given by Eq.~(\ref{V}) with $\alpha$ given by Eq.~(\ref{alpha}) have been conducted \cite{Fischbach Book,Franklin,Adelberger,Wagner,Will}, including the MICROSCOPE experiment \cite{Berge PRL,Berge arXiv,Fayet 2018,Fayet 2019}.  With few exceptions (e.g., \cite{Thieberger}), no  compelling evidence of a Yukawa interaction coupling to baryon number has been observed.

\begin{table}[t]
\caption{Values of the measured $\Delta\kappa$ with uncertainties $\sigma$ from the EPF experiment with $\Delta(B/\mu)$ for the pairs of materials in Ref.~ \cite{Fischbach PRL}.  Here the  column ``inflated $\sigma$'' contains the uncertainties needed for the EPF results to become consistent with $\Delta\kappa = 0$. For a description of the data point  Ag-Fe-SO$_4$, see Ref.~\cite{Fischbach PRL}. }
\begin{center}
\footnotesize
\begin{tabular}{|l|r|r|r|r|}
\hline 
Materials              		& $10^{8}\Delta\kappa$ & $\sigma$ & inflated $\sigma$ & $10^{3}\Delta(B/\mu)$ \\ \hline
Cu-Pt                  		& 0.4                	& 0.2       & 0.6        	& 0.94     \\ \hline
Magnalium-Pt        		& 0.4                	& 0.1       & 0.5        	& 0.50      \\ \hline
Ag-Fe-SO$_4$       	& 0.0                	& 0.2       & 0.2         	& 0.00       \\ \hline
Asbestos-Cu    			& $-0.3$       	& 0.2       & $-$0.1     & $-$0.74    \\ \hline
CuSO$_4$$\cdot$5H$_2$O-Cu	& $-0.5$      	& 0.2       & $-$0.3     & $-$0.84    \\ \hline
CuSO$_4$(solution)-Cu 	& $-0.7 $       	& 0.2       & $-$0.5   	& $-1.42$   \\ \hline
H$_2$O-Cu              		& $-$1.0           	& 0.2       & $-$0.8    	& $-$1.71   \\ \hline
\end{tabular}
\end{center}
\label{EPF data table}
\end{table}

\section{The E\"otv\"os Paradox}

We can now summarize the present situation regarding the EPF experiment with the following three statements \cite{Fischbach PoS}:
\begin{enumerate}
\item There is no compelling evidence of errors in the methods or results of the EPF experiments.
\item There is no compelling evidence to suggest a mistake in the reanalysis of the EPF results by Fischbach et al., which leads to the observed dependence of $\Delta\kappa$ on $\Delta(B/\mu)$ shown in Fig.~\ref{Original EPF graph}.
\item There is no compelling reason to question the results of the many experiments which show no evidence of a Yukawa fifth force coupling to baryon number.
\end{enumerate}
How can all of these seemingly mutually contradictory statements be  true?  The remainder of this paper  will explore possible answers to this ``E\"otv\"os Paradox.''

\section{The E\"otv\"os Pattern}

A solution to this paradox was suggested by Fischbach et al. with the important observation that one should focus on the  {\em pattern} revealed in the EPF data  as unambiguous evidence for a new interaction coupling linearly to $B$ \cite{Fischbach Respond}.  To understand their argument, we begin by reexamining Fig.~\ref{Original EPF graph}, the pattern observed when the values of $\Delta\kappa$ for the various combinations of materials used in the EPF experiment were plotted versus $\Delta(B/\mu)$ as in Ref.~\cite{Fischbach PRL}.   (Several other materials were used, but their baryon numbers were more difficult to characterize.   A more detailed discussion can be found in Ref.~\cite{AoP}.)  We note several features which will be important for our later discussion.  First, the data follow a linear pattern,
\begin{equation}
\Delta\kappa = a\, [\Delta(B/\mu)] + b,
\end{equation}
where a weighted least-squares fit finds a slope $a$ and intercept $b$ given by
\begin{align}
a & = (5.65 \pm 0.71) \times 10^{-6}, \\
b & = (4.83 \pm 6.44) \times 10^{-10},
\end{align}
with $\chi^2$ = 2.1 (5 degrees of freedom) \cite{Fischbach PRL}.  We observe two things from the E\"{o}tv\"{o}s pattern:  First, the slope deviates from the null hypothesis $a = 0$ by approximately $8\sigma$.  Second, the intercept $b$ is consistent with zero.  The latter observation will become important later, and has  not been emphasized sufficiently in most previous discussions of these results.

\begin{figure}[t]
\includegraphics[width=\columnwidth]{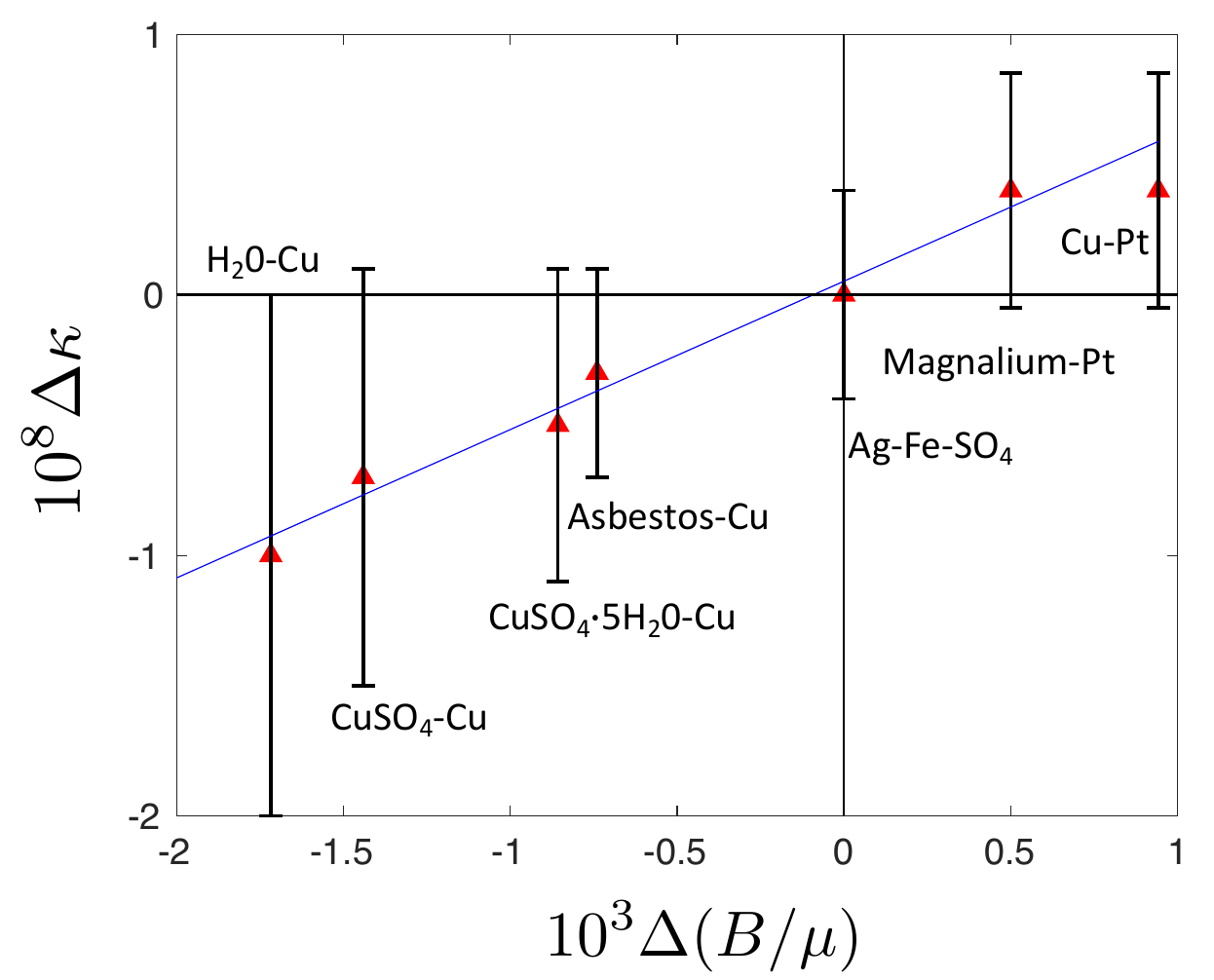}
\caption{The original EPF data given the seven pairs of materials in Table 1 of Ref [1] with inflated uncertainties such that a null effect is obtained for each individual point.}
\label{Inflated EPF graph}
\end{figure}

One might argue that the fact that most of the EPF data points are not consistent with the WEP at the $1\sigma$ level is an indication that the estimated uncertainties are too small.  If we artificially inflate the uncertainties $\sigma_{i}$ to bring the data into agreement with $\Delta\kappa = 0$, as in Table~\ref{EPF data table} and Fig.~\ref{Inflated EPF graph}, we find that the linear fit does not change significantly.  The slope and intercept  of the data with the inflated uncertainties are
\begin{align}
a_{\rm inflated} & = (5.09 \pm 2.44) \times 10^{-6}, \\
b_{\rm inflated} & = (0.20 \pm 1.93) \times 10^{-9}.
\end{align}
The slope remains nonzero, although its significance is reduced from 8$\sigma$ \space to 2.1$\sigma$.  Similarly, the intercept remains consistent with zero.  More importantly, though,   we find that  $\chi^2$ = 0.19 (5 degrees of freedom) with the modified data, which is artificially low, indicating that  the inflated uncertainties are in fact too large.

It is interesting to note that since the EPF experiment, we are not aware of any experiment of comparable or greater precision that  has compared the accelerations of three or more  pairs of materials of different composition at the same location in the same apparatus, which  would thus have been able to validate the E\"otv\"os pattern.   Some experiments, such as the floating ball experiments of Thieberger  \cite{Thieberger}, or Bizzeti and Bizzeti-Sona \cite{Bizzeti}, are \emph{sui generis} in the sense that they would be difficult to repeat with significantly different test samples. However, the designs of some other experiments could in principle allow for direct comparisons of the accelerations of different pairs of materials.   This would include the free-fall experiments of Niebauer, et al.  \cite{Niebauer}, Cavasinni, et. al. \cite{Cavasinni}, Kuroda and Mio \cite{Kuroda}, as well as the torsion balance experiments of the E\"{o}t-Wash collaboration \cite{Wagner}.  

\begin{figure}[t]
\includegraphics[width=\columnwidth]{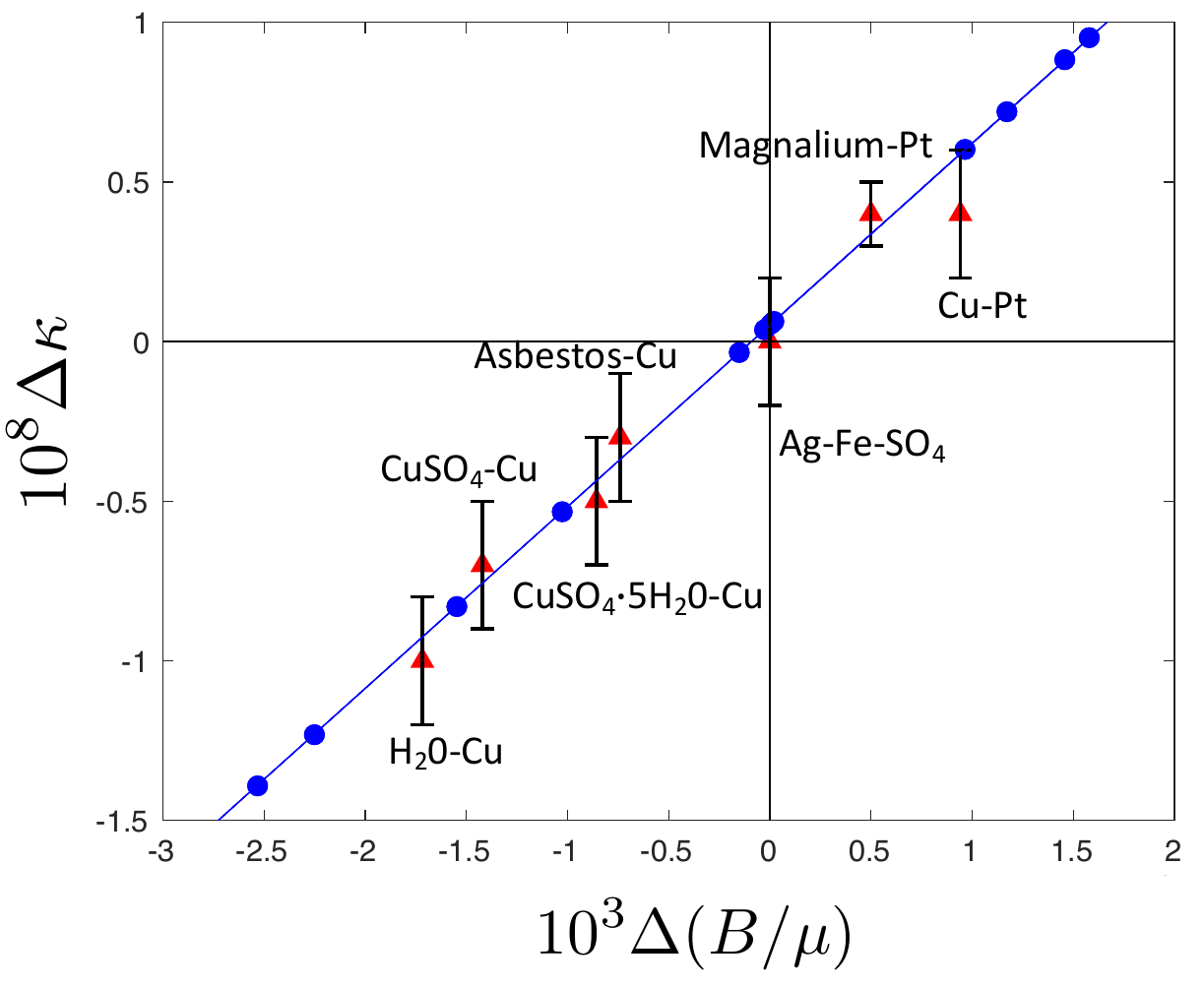}
\caption{A graph of the original EPF data (red triangles) along with hypothetical results that might have been obtained by EPF if they used the pairs of materials given in Table~\ref{New pairs table} (blue dots).}
\label{Hypothetical EPF graph}
\end{figure}

The latter group appears to be very well-suited to carry out an appropriate series of experiments with 3 or more independent pairs of samples. In Ref.~\cite{Wagner} the authors specifically compared the relative accelerations to Be-Ti, and of Be-Al, and hence adding one or more additional pairs seems well within the reach of this collaboration. In Table~\ref{New pairs table} we list the $\Delta(B/\mu)$ values for 12 representative pairs of samples which vary in magnitude over a range of approximately 400. (The data for Table~\ref{New pairs table} are obtained from Table 2.1 of Ref. \cite{Fischbach Book}). For illustrative purposes, we have created  in Fig.~\ref{Hypothetical EPF graph} a hypothetical plot  if these pairs of materials followed the same trend as the actual pairs studied by EPF.  It would thus appear that by using these (or similar) test samples, a definitive test of the composition-dependence of a fifth force can be carried out. Although in principle the source of a generic fifth force could be any ``charge'', the implication of Ref. \cite{Fischbach PRL} is that the EPF results arise specifically from a coupling of the baryon number, $B$, of a source to the baryon numbers $B_i$ of the test samples, as in Eq. (\ref{Delta kappa 5}). Although EPF give no explanations for their choice of test masses, it appears that their samples were selected to represent a fairly wide range of commonly available substances. However, given the suggestion in \cite{Fischbach PRL} of a possible coupling to $B$,  it is remarkable that the choices that EPF actually made constituted an almost ideal set of samples for present purposes:
\begin{table}[t]
\caption{12 suggested pairs of materials for further experimental searches for composition-dependent effects.  Data for $\Delta(B/\mu)$ values are from Ref.~\cite{Fischbach Book}, and  the values of $\Delta\kappa_{\rm hyp}$ are hypothetical values of $\Delta\kappa$ that would have been obtained from the EPF experiment had they been used, and if they followed the trend established by the actual samples used in the experiment.}
\begin{center}
%\footnotesize
\begin{tabular}{|l|r|r|}
\hline
Suggested Pairs     & $10^{3}\Delta(B/\mu)$ 	& $10^{8}\Delta\kappa_{\rm hyp}$ \\ \hline
Co-Cr    	& 0.006       & 0.056          		  \\ \hline
Be-Al   	& 0.006        & 0.056         	 \\ \hline
Be-Ti      	&0.020       & 0.064     		 \\ \hline
Fe-Cr  	 & 0.020       & 0.064         		   \\ \hline
Fe-Co   	& 0.026        & 0.138           		 \\ \hline
W-Pb   	& 0.151         & 0.138          		  \\ \hline
Pt-Cr   	& 0.965         & 0.602           		  \\ \hline
Pb-Co  	& 1.03        & 0.637           		  \\ \hline
C-Fe     	& 1.17         & 0.720          		  \\ \hline
Bi-Be       	& 1.46        & 0.883          		   \\ \hline
Pt-Be       	& 1.55       & 0.934        		  \\ \hline
Os-Be 	& 1.58          & 0.952           		 \\ \hline
Ag-Be     	& 2.25        & 1.34          	 \\ \hline
Fe-Be    	& 2.53         & 1.50         		  \\ \hline
\end{tabular}
\end{center}
\label{New pairs table}
\end{table}

\begin{enumerate}

\item  The samples incorporate 10 elements from various parts of the periodic table: H ($Z = 1$), C ($Z=6$), O ($Z=8$), Mg ($Z=12$), Al ($Z=13$), Si ($Z=14$), Fe ($Z=26$), Cu ($Z =29$), Ag ($Z=47$), and Pt ($Z=78$).

\item  The widespread use of Cu in 7 of their samples was especially fortunate given that Cu is near the peak in a plot of $B/\mu$ vs. $Z$.  This leads to relatively large values of $\Delta(B/\mu)$, thus increasing the sensitivity of the EPF experiment to a potential fifth force.

\item  The datum we denote as Ag-Fe-SO$_{4}$ is a comparison of the chemicals before and after a purely chemical reaction, and so fixes a point at the origin of the $\Delta \kappa$-$\Delta(B/\mu)$ plane.  As observed previously, the fact that the plot of $\Delta\kappa$ versus $\Delta(B/\mu)$ passes through the origin is an additional composition-dependent effect that is an unambiguous prediction of an interaction coupling linearly to baryon number (or hypercharge).

\item  The fact that some of the EPF samples contain the same elements, but in different proportions and combinations, makes it even less likely that the $\Delta\kappa$-$\Delta(B/\mu)$ correlation could have arisen by a statistical fluctuation.

\end{enumerate}

\section{Producing the  E\"otv\"os Pattern}

If we assume that the E\"otv\"os pattern is not the result of a statistical accident, but due to possibly new physics, what does it reveal about its cause?  The most important clues are provided by the linear fit to the pattern: The slope is positive and the intercept is zero.  Both of these features can be explained by a model in which each test body is subject to a force that is proportion to the sample's baryon number.  If $m_{i}$ is the mass of the $i$th test body and $B_{i}$ is its baryon number, then the acceleration of the test body is given by
\begin{equation}
\vec{a}_{i} = \vec{g} + \frac{B_{i}}{m_{i}}\vec{F}_{B} = \vec{g} + \left(\frac{B}{\mu}\right)_{i} \frac{\vec{F}_{B}}{m_{H}},
\label{a i}
\end{equation}
where $\vec{F}_{B}$ is the baryonic force per  baryon number.   Therefore, the difference in accelerations of two test bodies  relative to their gravitational accelerations is
\begin{equation}
\frac{\Delta\vec{a}}{g}=  \Delta \left(\frac{B}{\mu}\right)\vec{\cal F},
\label{a/g}
\end{equation}
where $\vec{\cal F} \equiv  \vec{F}_{B}/m_{H}g$ is the dimensionless force.  It is clear that Eq.~(\ref{a/g}) will produce the E\"otv\"os pattern: If $\Delta \kappa$ obtained from Eq.~(\ref{a/g}) is plotted versus $\Delta(B/\mu)$, the result will be a line with a slope $|\vec{\cal F}|$ {\em and} a vanishing intercept.  While most attention in the literature has been focused on the slope, this model also predicts that the intercept is zero.  There is obviously no baryonic force if a test body has a vanishing baryon number.  These two features of the E\"otv\"os pattern---a positive slope and vanishing intercept---are difficult to achieve from a random set of data.

While the E\"otv\"os pattern tells us much about what may be producing it, determining ${\cal F}$ requires a more specific  theoretical model.  For example,
 the original fifth force hypothesis, which proposed that the E\"otv\"os pattern was due to a Yukawa force coupling to baryon number,  gives
\begin{equation}
{\cal F} = {\cal F}_{5} =  \frac{f^2\epsilon(R/\lambda)}{G m^2_H} \left(\frac{B_\oplus}{\mu_\oplus}\right).
\end{equation}
This case also assumes that the source of the fifth force is the Earth, and that $\lambda$ has sufficient range that the force affects the EPF apparatus.  This example reveals that ${\cal F}$ is determined not only by the interaction model, but also by the source and the experimental setup.   Therefore, it may be difficult to compare the values of ${\cal F}$ obtained from different experiments without knowing the underlying theory.  

 There is currently much theoretical and experimental interest in the modification of gravity so that baryon number and/or quark content plays a role in observed interactions.   For example, a violation of the equivalence principle can arise from the exchange of a scalar boson, such as a dilaton.  However, in some of these theories, the leading order dependence of the force would be $B^{1/3}$, not $B$ \cite{Damour} and so will not produce the E\"otv\"os pattern.   
 While it is simplest to think of $\vec{\cal F}$ as a force field, as in the original fifth force hypothesis, there may be other possibilities. An alternative example would be a force due to elastic scattering of dark matter (DM) particles (e.g., WIMPs).   If the DM particles scatter incoherently off nuclei, the force would be proportional to $B$; if  DM scatters coherently, then the force would be proportional to $B^{2}$, and would not lead to the E\"otv\"os pattern.  Hence, the observation of the E\"otv\"os pattern can discriminate among theories which couple to baryon number in different ways.

\section{Reconciling with Other Experiments}

Since the discovery of the E\"otv\"os pattern in the reanalysis of the EPF data in 1986, a substantial experimental effort  has been undertaken to search for the fifth force and other putative forces predicted by string theory and other extensions of the Standard Model \cite{Tino,Franklin,Fischbach Book,Adelberger,Wagner,Will,Fischbach Metrologia}.  Yet, none of these experiments, some of which have significantly higher precision than the EPF experiment,  has found compelling evidence of new physics.  Can this be consistent with E\"otv\"os pattern? We consider several possibilities.

We begin by recognizing that experiments searching for new forces are generally  separated into two categories: composition-independent and composition-dependent experiments \cite{Fischbach Book}.  Examples of composition-independent  experiments are those which search for deviations from the inverse-square-law of Newtonian gravity.  Tests of the WEP, such as the EPF experiment, which are searching for acceleration differences between bodies of different composition, are examples of composition-dependent experiments. While an interaction which might be causing the E\"otv\"os pattern could appear in a composition-independent experiment, these experiments are not-optimized to search for a force which depends on baryon number.  Also, the E\"otv\"os pattern reveals little spatial information on what might be causing it and so it is not clear if  ${\cal F}$ of constant magnitude on the apparatus (e.g., a DM wind force) would be seen in an experiment optimized to search for deviations from the inverse-square-law.

While modern tests of the WEP are obviously composition-dependent, it is striking to note that, unlike the EPF experiment, they typically involve comparing one or two pairs of test bodies and so cannot reveal the E\"otv\"os pattern.  Of course, one can argue that this should not matter if these experiments have much higher precision than the EPF experiment.  If, however, the new interaction is coming from an unanticipated source, it may be zeroed out in the calibration as an unwanted background effect.  Similarly, experiments are optimized to be most sensitive to a specific sought-after signal.  This means that they are less sensitive to a signal from an unexpected source.  For example, Mueterthies has shown that  the EPF experiment is more sensitive to certain types of forces than the E\"ot-Wash WEP experiments (and vice versa), even though both use torsion balances \cite{Mueterthies arXiv,Mueterthies thesis}.    It is likely a similar analysis applies to comparing the MICROSCOPE experiment to the EPF experiment because of the significant difference in their designs.  Furthermore, the EPF experiment is sensitive to shorter-ranged interactions than MICROSCOPE, which uses the Earth as a source.  This is a  result of the difference in the designs of their setups, and the manner in which $\Delta\kappa$ is extracted from the measurements.  A similar argument can be made when comparing results from experiments using very different designs from the EPF experiment (e.g., free-fall, quantum interferometry, etc.).

\section{Discussion}

Nearly 35 years after its discovery, the origin of the pattern formed by the sloping line shown in  Fig.~\ref{Original EPF graph} of $\Delta \kappa$ versus $\Delta(B/\mu)$ obtained by the reanalysis of the EPF experiment remains a mystery.  While the original hypothesis of a Yukawa force coupling to baryon number (the fifth force), has since been excluded by many experiments, we have shown that the observed pattern can be explained by any force on the test bodies that is proportional to baryon number.   Therefore, it remains possible that the interaction responsible for the pattern may have remained undetected by experiments optimized to search for other types of forces.   Furthermore, recent composition-dependent experiments have used test bodies with insufficient variation in composition to exhibit the pattern.  

Since the actual source of the pattern remains unknown, the most model-independent approach to determine its cause would be to design composition-dependent experiments, such as EPF and earlier tests of the WEP, which use test bodies with the widest possible variation in baryon number, including pairs of samples with $\Delta(B/\mu) = 0$.  It is unlikely that confounding systematic effects in these experiments would coincidently scale with $B$.    To reduce this possibility,  the greatest number of varying pairs of samples is desired.   Since the effect on the apparatus may be dependent on the setup and local variables, the slope of the pattern may vary from experiment to experiment, but a non-zero slope with a vanishing intercept should be common to all.

\pagebreak

\end{document}